\documentclass[runningheads]{llncs}
\usepackage{graphicx}
\usepackage{cite}
\usepackage{hyperref}

\makeatletter
\newcommand{\printfnsymbol}[1]{%
  \textsuperscript{\@fnsymbol{#1}}%
}
\makeatother
\begin{document}
\title{Cross-Modality Domain Adaptation for Vestibular Schwannoma and Cochlea Segmentation}

\author{Han Liu\thanks{equal contribution} \and
Yubo Fan\printfnsymbol{1} \and
Can Cui \and
Dingjie Su \and
Andrew McNeil \and
Benoit M. Dawant}

\authorrunning{H. Liu et al.}
%
\institute{Vanderbilt University, Nashville TN 37235, USA\\
\email{han.liu@vanderbilt.edu}}
\maketitle              
\begin{abstract}
Automatic methods to segment the vestibular schwannoma (VS) tumors and the cochlea from magnetic resonance imaging (MRI) are critical to VS treatment planning. Although supervised methods have achieved satisfactory performance in VS segmentation, they require full annotations by experts, which is laborious and time-consuming. In this work, we aim to tackle the VS and cochlea segmentation problem in an unsupervised domain adaptation setting. Our proposed method leverages both the image-level domain alignment to minimize the domain divergence and semi-supervised training to further boost the performance. Furthermore, we propose to fuse the labels predicted from multiple models via noisy label correction. Our results on the challenge validation leaderboard showed that our unsupervised method has achieved promising VS and cochlea segmentation performance with mean dice score of 0.8261 ± 0.0416; The mean dice value for the tumor is 0.8302 ± 0.0772. This is comparable to the weakly-supervised based method.

\end{abstract}

\section{Introduction}
Vestibular schwannoma (VS) is a benign tumor that arises from the Schwann cells of the vestibular nerve, which connects the brain and the inner ear. To facilitate the follow-up and treatment planning of VS,  automatic methods to segment the VS tumors and the cochlea from magnetic resonance imaging (MRI) are proposed \cite{1}. While the most commonly used modality for VS segmentation is contrast-enhanced T1 (ceT1), high-resolution T2 (hrT2) imaging has been demonstrated to be a possible alternative with less risk and lower cost.\cite{2}. 

Supervised segmentation methods have shown the good performance in VS segmentation \cite{supervised}, but they require to fully annotate image data which may not be an option in practice. Weakly-supervised methods require less annotation efforts, such as scribbles and bounding boxes, and sometimes they even achieve comparable performance to the supervised ones \cite{weakly}. In this work, we aim at segmenting the VS tumor and the cochlea in hrT2 without any hrT2 annotations during training. Specifically, we are provided with a dataset consisting of ceT1 images and hrT2 images, but only the ceT1 images have the segmentation labels. We consider the problem as an unsupervised domain adaptation (UDA) problem. There are mainly two types of methods to tackle the UDA problem, domain alignment and techniques based on semi-supervised learning (SSL). Domain alignment focuses on reducing the distribution discrepancy by optimizing some divergence metric \cite{31,36} or via adversarial learning \cite{27, DANN}. On the other hand, self-training \cite{selftrain}, mean teacher \cite{mt}, and other SSL-basd techniques also offer competitive performance. In this work, we focus on exploring methods that combines image-level domain alignment and SSL for UDA.

\section{Methods}
\subsection{Problem formulation}
For an unsupervised domain adaptation problem, we have access to a source domain $D^{S}=\{(x^{s}_{i}, y^{s}_{i})|i=1,2,\cdots,n_{s}\}$, and a target domain $D^{T}=\{x^{t}_{j}|j=1,2,\cdots,n_{t}\}$, where $Y^{S}$ and $Y^{T}$ share the same $K$ classes. In our case, source and target domains correspond to ceT1 and hrT2 respectively and $K=3$ representing background, VS and cochlea. We aim to train a segmentation network $F_{t}$ that learns the knowledge from the source domain and is capable to achieve robust and accurate segmentation performance on the target domain, without accessing the target domain labels $Y^{T}$. 

\subsection{Image-level Domain Alignment}
Image-level domain alignment is a simple but effective method to tackle UDA problem by reducing the distribution mismatch at the image-level, i.e., pseudo image synthesis. Here, we propose to train the segmentation model $F_t$ with the pseudo target domain images $\tilde{X}^{T}$, which are generated by unpaired image-to-image translation. We explored both end-to-end training and two-stage training. For end-to-end training, we rely on the Contrastive Unpaired Translation (CUT) \cite{cut} as backbone for image synthesis and add an extra segmentation module $F_t$ on top of the synthesized images. This method will be referred as \textbf{CutSeg}. We select CUT for unpaired image-to-image translation because it can be trained faster and is less memory-intensive, allowing more flexibility when adding the 3D CNN-based segmentation module. During training, we first train the CUT model alone till it achieves reasonable synthesis performance. Then we train the CutSeg end-to-end with the CUT subnetwork initialized with the pre-trained weights and the segmentation module trained from scratch. For two-stage training, we used the CycleGAN \cite{cyclegan} to generate pseudo hrT2 images $\tilde{X}^{T}$. To improve the data diversity, we trained both 2D and 3D CycleGANs and collected pseudo images from different epochs. Lastly, we trained a segmentation module $F_t$ using $\tilde{X}^{T}$.

\subsection{Semi-supervised Training}
Though image-level domain alignment can minimize the domain divergence, the unlabeled target domain images $X^{T}$ are not directly involved in training the segmentation model $F_{t}$. To overcome this limitation, we propose to adapt a semi-supervised learning method named Mean Teacher (MT) \cite{cv_mt} to make better use of $X^{T}$. Specifically, a student model along with a teacher model with the same network architecture are created and both models are initialized with the best model weights obtained from Section 2.2. In our semi-supervised setting, the labeled images are the pseudo hrT2 images while the unlabeled images are the real hrT2 images. During training, the labeled pseudo images are fed to student model and the segmentation loss $L_{seg}$ is computed in a supervised manner. For unlabeled images, we first augment the same image twice with different intensity transformation parameters. The augmented images are then fed to the student model and the teacher model separately and a consistency loss $L_{con}$ is computed. Both $L_{seg}$ and $L_{con}$ are used to update the weights of the student model and the weights of teacher model are updated as an exponential moving average of the student weights. As suggested in \cite{cv_mt}, the teacher prediction is more likely to be correct at the end of the training and thus the teacher model is taken as our final $F_{t}$.

\subsection{Noisy Label Correction as Label Fusion}
In this challenge, we have obtained three models that were trained from different strategies and each model alone has achieved satisfactory result on the validation leaderboard. The first model is obtained by two-stage training using the pseudo images from 2D CycleGAN, followed by semi-supervised training by MT. The second model is fine-tuned based on the first model using the pseudo images from 3D CycleGAN. The third model is a CutSeg model. Training details can be found in Section 3.1.

\begin{figure}[h]
\includegraphics[width=1\columnwidth]{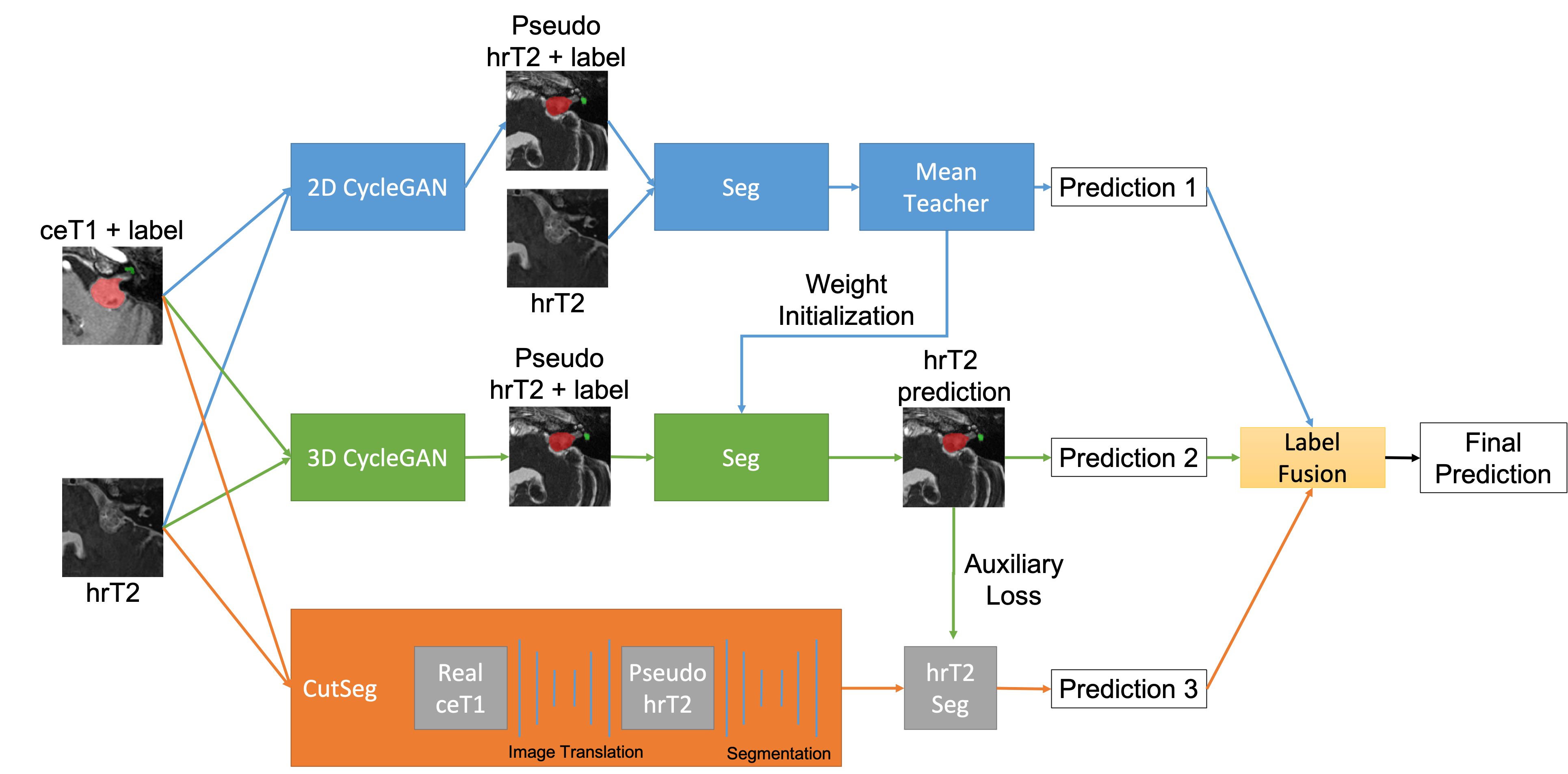}
\centering
\caption{The schematic diagram of our proposed method} \label{fig1}
\end{figure} 

Empirically, ensembles tend to yield better predictive performance when there is a significant diversity among the models. Here, we propose to fuse the labels from different models by treating the label fusion task as a noisy label correction problem. We adapted a confident learning method called \textbf{cleanlab} \cite{cleanlab} which provides exact noise estimation and label error finding. Note that we use cleanlab to directly fuse labels at the inference phase rather than update the pseudo labels iteratively during training. Specifically, we first obtain the softmax outputs of two models and convert one output to a one-hot encoded label mask. The one-hot encoded mask was considered as a 'noisy label' and was corrected by the softmax outputs from the other model. Once the labels from the first two models were fused, the fused labels are treated as noisy labels and fused again with the softmax outputs from the remaining model. The labels fused from three models are used as our final predictions.

\section{Experiments and Results}
\subsection{Data and Implementation.}
The dataset was released by the MICCAI challenge CrossMoDA 2021. All images were obtained on a 32-channel Siemens Avanto 1.5T scanner using a Siemens single-channel head coil \cite{data}. ceT1 imaging was performed with an MPRAGE sequence with in-plane resolution of 0.4×0.4 mm, in-plane matrix of 512×512, and slice thickness of 1.0 to 1.5 mm. For hrT2 images, imaging was performed with a 3D CISS or FIESTA sequence in-plane resolution of 0.5×0.5 mm, in-plane matrix of 384×384 or 448×448, and slice thickness of 1.0 to 1.5 mm. The VS and cochleas were manually segmented in consensus by the treating neurosurgeon and physicist using both the ceT1 and hrT2 images. We randomly split the images into 185 and 25 for training and validation respectively. Since the field of views (FoV) of the source and target domain images vary significantly, we crop each image into a cubic box, or ROI, using single-atlas registration \cite{ants}. The ROI on the atlas image is manually cropped around the right side of the brain. To obtain the ROI on the left side, we flip the volume left-to-right before performing registration. 

For preprocessing, in two-stage training, we resample the images to the most common spacing in the target domain, i.e., (0.46875, 0.468975, 1.5) and normalize the intensity to [0, 1]. In end-to-end training, we first train a CutSeg models for 82 epochs with an auxiliary consistency loss, which is a Mean Absolute Error (MAE) loss between the segmentation result of the real hrT2 image and the prediction from the two-stage training model. Then the CutSeg model was fine-tuned on the hrT2 images with in-plane resolution higher than 0.5 mm. During inference, the fine-tuned CutSeg model was used to make inference for the testing images with in-plane resolution above 0.5 mm. For the segmentation module, we adapted the model architecture from \cite{weakly} and used dice + cross-entropy loss for training. For post-processing, we first reduce the false positive VS prediction by removing the isolated components whose center is greater than the adjacent cochlea center by 15 voxels along z-axis. Then we take the largest connected components for both VS and cochlea within each ROI.

For training, we used Adam optimizer with weight decay $10^{-4}$ and batch size $1$. The learning rates were initialized to $5\times10^{-4}$, $5\times10^{-5}$ and $2\times10^{-4}$ for two-stage training, MT and CutSeg respectively. The hyperparameters were determined by grid-search within the range of $10^{-2}$ to $10^{-6}$. The best hyperparameters were selected based on the segmentation performance on our own validation set. The CNNs were implemented in PyTorch \cite{pytorch} and MONAI on a Ubuntu desktop with an NVIDIA RTX 2080 Ti GPU. For quantitative evaluation, we measured the Dice score and average symmetric surface distance (ASSD) between segmentation results and the ground truth.

\subsection{Experimental Results}
\begin{figure}[t]
\includegraphics[width=1\columnwidth]{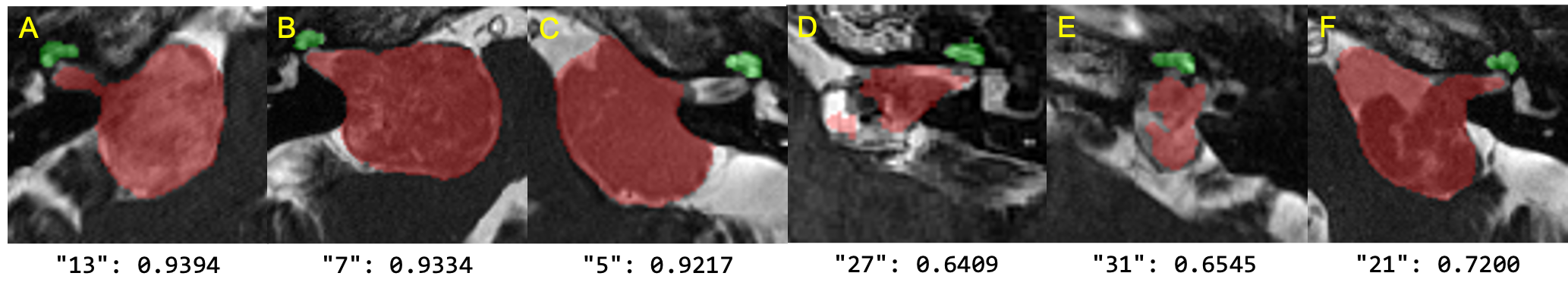}
\centering
\caption{Quantitative results. A to C and D to F show the best and worst VS segmentation results. The image ID and the corresponding dice score are displayed.} \label{fig2}
\end{figure} 

The following table shows the evaluation metrics of our proposed method on the validation leaderboard. Due to the page length limit, the ablation studies and other explored methods are not included in this paper. 

\begin{table}[]
\caption{Quantitative results on validation leaderboard}
\centering
\begin{tabular}{|l|l|l|}
\hline
        & Dice            & ASSD            \\ \hline
VS      & 0.8302 ± 0.0772 & 0.5686 ± 0.2675 \\ \hline
Cochlea & 0.8220 ± 0.0310 & 0.1829 ± 0.0476 \\ \hline
\end{tabular}
\end{table}

\section{Conclusion}
In this work, we exploited the image-level domain alignment and semi-supervised training to tackle the unsupervised domain adaptation segmentation problem. According to the validation leaderboard, our unsupervised method has achieved a segmentation performance that is comparable to the performance of the weakly-supervised method, demonstrating its effectiveness.

%
%
%

\end{document}